\documentclass[aps,prd,twocolumn,groupedaddress,showpacs,preprintnumbers,amsmath,amssymb,floatfix]{revtex4}
\usepackage{graphicx}% Include figure files
\usepackage{subfig}
\usepackage{dcolumn}% Align table columns on decimal point
\usepackage{float}
\usepackage{bm}% bold math
\usepackage{enumerate}
\usepackage[english]{babel}
\usepackage{amsfonts}
\usepackage{amsthm}
\usepackage{amssymb}
\usepackage{calc}
\usepackage{ifthen}
\usepackage{color}
\usepackage{hyperref}
\usepackage{bibcommands}

\renewcommand{\Im}{\mathrm{Im}}

\newcommand{\GL}{Ginzburg-Landau }

\DeclareMathOperator{\defin}{\equiv}

\newcommand{\numset}[1]{\mathbb{#1}}

\DeclareMathOperator{\union}{\cup}

\newcommand{\integers}{\numset{Z}}
\newcommand{\reals}{\numset{R}}
\newcommand{\complexes}{\numset{C}}
\newcommand{\sphere}{\mathsf{S}}

\newcommand{\abs}[1]{\left\lvert#1\right\rvert}

\newcommand{\norm}[1]{\lVert#1\rVert}

\providecommand{\scalarprod}[2]{#1 \cdot #2} % same as innerprod, but without ^T notation 

\newcommand{\crossprod}{\times}
\DeclareMathOperator{\combmap}{\circ}
\newcommand{\conjugate}[1]{\overline{#1}}
\newcommand{\adjoint}[1]{#1^\dagger} % ordinary (hermitian) adjoint of #1
\newcommand{\extder}{\text{d}} % exterior derivative

\DeclareMathOperator{\isom}{\cong}

\newcommand{\density}[1]{\mathcal{#1}}

\newcommand{\half}{\tfrac{1}{2}}

\renewcommand{\vec}[1]{\boldsymbol{#1}}

\theoremstyle{plain}
\theoremstyle{definition}
\theoremstyle{remark}
\begin{document}

\title{Knot solitons in modified Ginzburg-Landau model}

\author{Juha J\"aykk\"a}
\email{juhaj@iki.fi}
\affiliation{
School of Mathematics,
University of Leeds,
LS2~9JT,
United Kingdom
}
\author{Joonatan Palmu}
\email{jjmpal@utu.fi}
\affiliation{
Department of Physics and Astronomy,
University of Turku,
FI-20014 Turku, Finland
}

\date{\today}

\begin{abstract}
  We study a modified version of the Ginzburg-Landau model suggested by Ward and show that
  Hopfions exist in it as stable static solutions, for values of the Hopf invariant up to
  at least 7. We also find that their properties closely follow those of their
  counterparts in the Faddeev-Skyrme model. Finally, we lend support to Babaev's
  conjecture that longer core lengths yield more stable solitons and propose a possible
  mechanism for constructing Hopfions in pure Ginzburg-Landau model.
\end{abstract}

\maketitle

\section{\label{sec:intro}Introduction}

Topological solitons have long enjoyed widespread interest within many fields of physics,
including such seemingly distant subjects as cosmology, condensed matter physics and
particle physics. Of these, the most common example are perhaps the Abrikosov vortices in
type II superconductors. Therefore, it is crucial to understand the properties and
existence of topological solitons. The purpose of this work is to provide further
information about their presence in Ward's modified Ginzburg-Landau model.

While many topological solitons are pointlike or two-dimensional, there are also extended
three-dimensional topological solitons; one particular class of them is called Hopfions,
their name arising from the topological invariant associated with them, called the Hopf
invariant. The archetype model supporting topologically stable {\it closed} vortices, is
the Faddeev-Skyrme model \cite{Faddeev:1975tz, Faddeev:1979aa, Faddeev:1997zj,
  Battye:1998pe, Battye:1998zn, Hietarinta:1998kt, Hietarinta:2000ci, Hietarinta:2003vn,
  Adam:2006bw}. While this has been know for quite some time, it has recently become more
relevant with the discovery of topological insulators and the possibility of the existence
of Hopfions in them \cite{Moore:2008aa}. It is also known
\cite{Hindmarsh:1992yy,Babaev:2001zy} that that the Faddeev-Skyrme model can be embedded
into the Ginzburg-Landau model, giving rise to a conjecture in \cite{Babaev:2001zy} that
the two-component \GL model could, due to this embedding, also support the same
topological structures as the FS model does.

Previous research has not been able to reach a conclusion on the stability of Hopfions in
the \GL model. There has been one positive result in a restricted model \cite{Niemi:2000ny} but it has not
been confirmed in other investigations using the full model
\cite{Ward:2002vq,Jaykka:2006gf,Jaykka:2009rw}. This suggests that any Hopfions in the \GL
model are very hard to find, so studying closely related models is not only relevant in
itself, but may also provide clues as to how to find Hopfion in the \GL model. In the
modified \GL model, however, stable Hopfions have been discovered
\cite{Ward:2002vq,Jaykka:2009rw}, but these works have only explored the stability of
Hopfions at Hopf invariant one. We will show that Hopfions exist as local minimal energy
configurations in the modified model up to at least the value $7$ of the Hopf invariant
and that these share several important features with their counterparts in the
Faddeev-Skyrme model. Finally, we provide support for a conjecture by Babaev
\cite{2009PhRvB..79j4506B} that solitons with longer cores are more stable and propose a
possible way to construct Hopfions in the unmodified Ginzburg-Landau model.

\section{\label{sec:model}The model}

The static Abelian Higgs model with two charged Higgs bosons has the same mathematical
form as the \GL model with two flavors of Cooper pairs or superfluids. We will use the
following notations.  The indices run as follows: $j,k,l \in \{1,2,3\}$, $\alpha \in
\{1,2\}$, 
and the fields are $\Psi = (\psi_1 \quad \psi_2)^T$, $F_{jk} \equiv \partial_j A_k
- \partial_k A_j$, $B^j = \epsilon^{jkl}\partial_k A_l$ and the gauge-covariant derivative
has the form $D_j \equiv \partial_j - i g A_j$; for short, we will also write $\vec{D}
\equiv \nabla - i g \vec{A}$.  With these notations, the standard static energy density of
the two-component \GL model can be written as
\begin{align}
  \label{eq:EtcGL}
   {\density E} &=
   \half \norm{\vec{D} \psi_\alpha}^2 + V\bigl(\psi_{1},\psi_{2}\bigr) + \tfrac{1}{2}\norm{\vec{B}}^2,
\end{align}
where we have used SI units. The exact form of the potential depends on the physical
context, but for the purposes of this article, all that is required is that it maintains
the $SU(2)$ symmetry of the model and enforces the condition $\norm{\Psi}=$~constant~$\ne
0$ at some limit of the parameters (now $\eta$) of the potential. Here we have used
\begin{align}
  \label{eq:potential}
  V(\psi_{1},\psi_{2}\bigr) &= \half \eta (\abs{\Psi}^2-1)^2.
\end{align}

The embedding of Babaev {\it et~al.} \cite{Babaev:2001zy} is useful to demonstrate how
closed (or knotted) vortices might exist in \GL model. These will be defined in terms of
the fields $\psi_\alpha$, leaving the gauge field $\vec{A}$ free. The embedding requires
that $\norm{\Psi}>0$ everywhere, but his is not enough to reveal the closed vortices and
further conditions need to be met as we shall next describe. In all the topological
considerations that follow we assume that $\norm{\Psi}>0$ everywhere and $\Psi$ can be
thought of as normalized by $\Psi \to \Psi/\norm{\Psi}$, which implies that $\Psi \in
\sphere^3$.

Maps $\sphere^3 \to \complexes P^1 \isom \sphere^2$ fall into disjoint homotopy classes,
the elements of $\pi_3(\sphere^2)$ and it is well known that $\pi_3(\sphere^2) =
\integers$. Thus, for each map $\sphere^3 \to \complexes P^1$ we can assign an integer,
called the Hopf invariant, which tells us which element of $\pi_3(\sphere^2)$ that belongs
to. Similarly, maps $\sphere^3 \to \sphere^3$ belong to elements of
$\pi_3(\sphere^3)=\integers$; this integer is called the degree. Suppose one has two maps:
$\Psi:\sphere^3 \to \sphere^3$ and the Hopf map $h:\sphere^3 \to \complexes P^1 \isom
\sphere^2$. Then it can be shown that the Hopf invariant of $h \combmap \Psi$ equals the
degree of $\Psi$. Since we work with maps $\sphere^3 \to \sphere^3$, the relevant
topological invariant is the degree, but there is always an associated Hopf invariant,
$H$, which has the same value. Topological solitons, i.e., stable static solutions $\Psi$,
with an associated Hopf invariant, are called Hopfions and also knot solitons, due to
their general shape at higher values of $H$. Next, we will see how it is the presence of
the Hopf invariant, not the degree, which gives rise to closed vortices and knot solitons.

Without loss of generality, we can choose $\Psi_\infty = (1, 0)$. Now, the Hopf map takes
$\Psi_\infty \mapsto \phi_\infty \defin (0,0,1) \in \sphere^2$ and we define the
soliton core as the preimage $(h \combmap \Psi)^{-1}(-\phi_\infty)$. 

It is natural to ask whether one of these Hopfions is the global energy minimum for each
$H$ as is the case in the Faddeev-Skyrme model. The answer to the question is, in general,
negative. The fact that $\vec{A}$ is left free, means that there is no nontrivial topology
imposed on it and since in the vacuum of the \GL model $\vec{A}$ is pure gauge, the
magnetic field energy can vanish in all cases. This in turn means that there is no longer
a fourth-order derivative in the energy density \eqref{eq:EtcGL} and Derrick's theorem
\cite{Derrick:1964ww} states that no stable, topologically nontrivial solutions of the
field equations with nonzero energy exist. This fact has been observed many times, by
various authors, including \cite{2008PhR...468..101R, 2009PhRvB..79j4506B,
  Speight:2008na}. It has also been seen in numerical work, that the magnetic field energy
indeed does vanish and with it, the soliton itself \cite{Ward:2002vq, Jaykka:2006gf,
  Jaykka:2009rw}. 
%This is indeed tentatively confirmed by
%\cite{Jaykka:2006gf, Ward:2002vq, 2009PhRvB..79j4506B, Jaykka:2009rw} and also an rigorous
%analytical work \cite{Speight:2008na}. 
In short, the knot soliton can always be undone by
decreasing the magnetic field to zero and radially shrinking it. However, if the starting
point is chosen suitably, this process may involve temporarily increasing the energy of
the configuration. Thus, local minima may still exist. Indeed, they were shown to exist in
the modified model in \cite{Ward:2002vq, Jaykka:2009rw}, while in \cite{Jaykka:2006gf} the
collapse of the magnetic field described above was observed and no stable solutions were
found in pure \GL model.

The crucial ingredient to finding these local minima is to prevent the collapse of the
magnetic field in order to obtain stable topologically nontrivial configurations in the
model. There are physical arguments that suggest there might exist physical processes that
prevent the collapse, but here we follow the path set out by \cite{Ward:2002vq}, where the
energy density is modified by adding the term $\density{E}_{W}$ (we denote $\adjoint{\Psi} =
(\conjugate{\psi_1} \quad \conjugate{\psi_2})$):
\begin{align}
  \label{eq:EGLW}
  \begin{split}
  \density{E}_{GLW} &=
    \overbrace{\tfrac{1}{2} \norm{\vec{D} \Psi}^2}^{\equiv \density{E}_K} + 
    \overbrace{\tfrac{1}{2} \norm{\nabla \crossprod \vec{A}}^2}^{\equiv \density{E}_B} \\
    &\quad + \overbrace{\half \kappa^2 \norm{\adjoint{\Psi} \vec{D} \Psi}^2}^{\equiv \density{E}_W} +
    \overbrace{V\bigl(\psi_{1},\psi_{2}\bigr)}^{\equiv \density{E}_P} .
  \end{split}
  \intertext{Denoting for any subscript $z$: $E_z = \int \extder^3x \density{E}_{z}$ we finally have the
    total energy}
  \label{eq:EGLW_total}
  E_{GLW} &= E_K + E_W + E_B + E_P.
\end{align}
The extra term, when the parameters $\kappa,\eta \rightarrow \infty$, ensures that the
model becomes exactly the Faddeev-Skyrme model 
\begin{align}
  \label{eq:EFS}
  \density{E}_{FS} &= \tfrac{1}{8} \norm{\partial_k \phi}^2 + \tfrac{1}{16} \norm{\scalarprod{\phi}{\partial_j 
  \phi \crossprod \partial_k \phi}}^2 ,
\end{align}
where $\phi$ is a normalized real valued field $S^3 \to S^2$. Therefore the model supports
Hopfions, at least asymptotically. This limit of $\kappa \to \infty$ was apparently first
observed by Hindmarsh \cite{Hindmarsh:1992yy}, albeit in a slightly different context.

\section{\label{sec:numerical_methods}Numerical methods}

We will now turn to our numerical investigation. All the computations are done using the
methods and programs described in \cite{Jaykka:2009rw}. In brief, we have used
gauge-invariant discretization of the energy density with simple forward differencing
scheme for derivatives. We then compute the gradient of the discrete energy density with
respect to the fields $\Psi, \vec{A}$ and use a conjugate gradient algorithm to find a
local minimum from a given initial configuration. The initial configurations of $\Psi$ for
lower values of $H$ are constructed using the methods of \cite{Jaykka:2006gf}, that is,
denoting the coordinates of $\reals^3$ by $x_i$ and $r^2 = x_1^2 + x_2^2 + x_3^2$, we set
\begin{subequations}\label{eq:Psiwithsech}
  \begin{align}
    \psi_1\bigl(\vec{x}\bigr) &= \tfrac{\sqrt{(r^2-1)^2+4x_3^2}}{r^2+1}
    \Bigl(\tfrac{r^2-1 + 2ix_3}{\sqrt{(r^2-1)^2+4x_3^2}}\Bigr)^p,
    \\
    \psi_2\bigl(\vec{x}\bigr) &= \tfrac{2\sqrt{x_1^2+x_2^2}}{r^2+1}
    \Bigl(\tfrac{x_1+ix_2}{\sqrt{x_1^2+x_2^2}}\Bigr)^q.
  \end{align}
\end{subequations}
These satisfy $\norm{\Psi}>0$ everywhere and indeed $\norm{\Psi}=1$ everywhere. The
degree of this configuration is $\deg \Psi = pq$, so we can easily construct
configurations of any given degree -- or Hopf invariant -- when one thinks about the
embedding. For $H \ge 5$, the initial configurations described in section 3 of
\cite{Sutcliffe:2007ui} were used.

We also need an initial configuration for $\vec{A}$, which we take to be always
\begin{multline}
  \label{eq:initialA}
  A_i(x) = -\Im \biggl(
  -\conjugate{\psi_1\bigl(x\bigr)}\Bigl(\psi_1\bigl(x+h\hat{e}_i\bigr)-\psi_1\bigl(x\bigr)\Bigr)\\
  -\conjugate{\psi_2\bigl(x\bigr)}\Bigl(\psi_2\bigl(x+h\hat{e}_i\bigr)-\psi_2\bigl(x\bigr)\Bigr)
  \biggr)/h,
\end{multline}
where $h$ is the lattice constant. The choice of initial value for $\vec{A}$ is largely
irrelevant since it is has the same homotopy (trivial) in any case, but
Eq.~\eqref{eq:initialA} convergences much faster relative to the obvious initial
configuration $A_i=0$.

In order to save computer resources (time), we reused previously found minimum
configurations to look for new ones: once a stable minimum was found for some values of
$H,\kappa,\eta$, we sometimes used that solution as an initial configuration for some new
values of $\kappa,\eta$ (of course, we cannot change $H$ in this manner). This initial
configuration converges faster than a fresh configuration set up using
Eq.~\eqref{eq:initialA}, which we interpret to indicate that the old minimizer is in some
sense ``closer'' to the new solution than the fresh configuration.
Furthermore, it is often true, that such a reused minimizer converges to a stable solution
even when the fresh initial configuration does not, that is, the fresh configuration is
sometimes outside the attraction basin altogether.
This emphasizes the fact that we can never be certain that there is no stable solution for
a given pair $\kappa,\eta$ using our methods: the attraction basin could simply be so tiny
that constructing an initial configuration within it is nearly impossible without some
additional knowledge about its location within the configuration space. We will, however,
present strong evidence pointing to the nonexistence of stable solutions for certain
values of $H,\kappa,\eta$.

In modern numerical work, data post-processing is in an increasingly complex role. We will
now describe the data post-processing used in this work. All the post-processing is done
using the \textit{MayaVi Data Visualizer} \cite{ramachandran2010mayavi}, with some
extensions we have implemented ourselves in the python language. Even though MayaVi is
designed for visualization, it contains routines generally useful for post-processing,
such as interpolation, which we use heavily. For an isosurface plot we compute $\phi=h
\combmap \Psi$ and then simply ask MayaVi to find, using interpolation, the surface
satisfying the conditions for the isosurface in question.

Computing the core length is a more complex operation since interpolation cannot be
directly used to find the minimum values of data. However, we can work around this
limitation using the fact that $\norm{\phi}=1$. We use MayaVi's routines to first filter
out points where $\phi_3>0$, since we are not interested in these. Then we apply MayaVi's
contour finding routine two successive times to the remaining data: first, we find the
contour $\phi_1=0$, i.e. points where $\phi^2_2 + \phi^2_3=1$ and then from this data, we
find the contour $\phi_2=0$, i.e. points where $\phi^2_3=1$, but since we excluded points
where $\phi_3>0$, what is left is the core $\phi_3=-1$. MayaVi represents this as a
polygon, whose circumference we then compute to give the core length. The main advantage
of this method is that MayaVi is able to interpolate the contours at each stage, giving
smoother cores than would be possible with direct methods. The accuracy of this procedure
is very good: at a lattice with $60^3$ points and lattice constant $h=0.1$, the
interpolated core length has an error of less than $0.03\%$.

\begin{figure}[h]
  \centering
  \includegraphics[width=0.45\columnwidth]{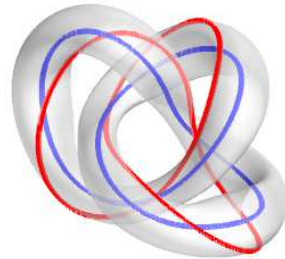}
  \caption{The isosurface of $\phi_3= -0.5$ for the local energy minimum at
$\kappa=\eta=10$. 
%The blue curve is the soliton core and the red curve
The solid curve inside the isosurface is the soliton core and the other solid curve is
another preimage, namely $\phi^{-1}(-0.87,0,-0.5)$.}
\label{fig:trefoiliso}
\end{figure}

As an example of previously described routine, we present the trefoil knot shaped local
minimum at $H=7$, $\kappa=\eta=10$ in Fig.~\ref{fig:trefoiliso}. The other preimage is
constructed in a similar fashion.

\section{\label{sec:results}Results}

We have investigated the properties of stable solutions in the modified \GL model for
$H\in [1,7]$. All numerical results have been obtained in a standardized lattice with
$360^3$ lattice points and lattice constant $h=1/12$. Since this is different from the
lattice of \cite{Jaykka:2009rw}, we also recomputed the $H=1$ case in the new standard
lattice in order to be able to compare the results for all values of $H$. Some checks were
made in lattices of $480^3$, $h=1/12$ and $180^3$, $h=1/6$ as well.

We chose to investigate the solutions (stable or otherwise) for each $H$ along two lines:
$\kappa=10$ and $\eta=10$. As expected, for both lines, we find that there is a limiting
value of $\kappa$ ($\eta$) below which no stable solutions can be found. These values
depend on $H$. For $H=1$ our results are in agreement with \cite{Ward:2002vq,
  Jaykka:2009rw} and stable solutions are also found for all investigated $H>1$. 

This procedure gives us two points on the stable/unstable boundary investigated in
\cite{Jaykka:2009rw}, and also information about how the stable solutions change with
changing $\kappa$ and $\eta$, which is not so easy to discern from solutions following the
actual stable/unstable boundary, where both $\kappa$ and $\eta$ are changing.

\subsection{\label{subsec:error_considerations}Error considerations}

Derrick's theorem \cite{Derrick:1964ww} implies that for any static solution of the field
equations which is stable against uniform scaling of space, a virial theorem must hold. In
this case, it takes the form $E_B = E_K + E_W + 3 E_P$. Obviously, for a numerical
approximation, the equation will only be satisfied to within some tolerance. This
tolerance would ideally be deduced from the discretization errors, effects of a finite
computational domain and the tolerance used to determine convergence, but this seems an
insurmountable task. Instead, we have computed the same solution with different lattices
(varying both size and density) to give us an estimate of the accuracy. This then gives us
an idea of the tolerance in the virial theorem which is achievable at a given
lattice. We use this tolerance  to estimate the errors in the results.

\begin{figure}[h]
  \centering
  \subfloat[][$\kappa=10$]{\includegraphics[width=0.45\columnwidth]{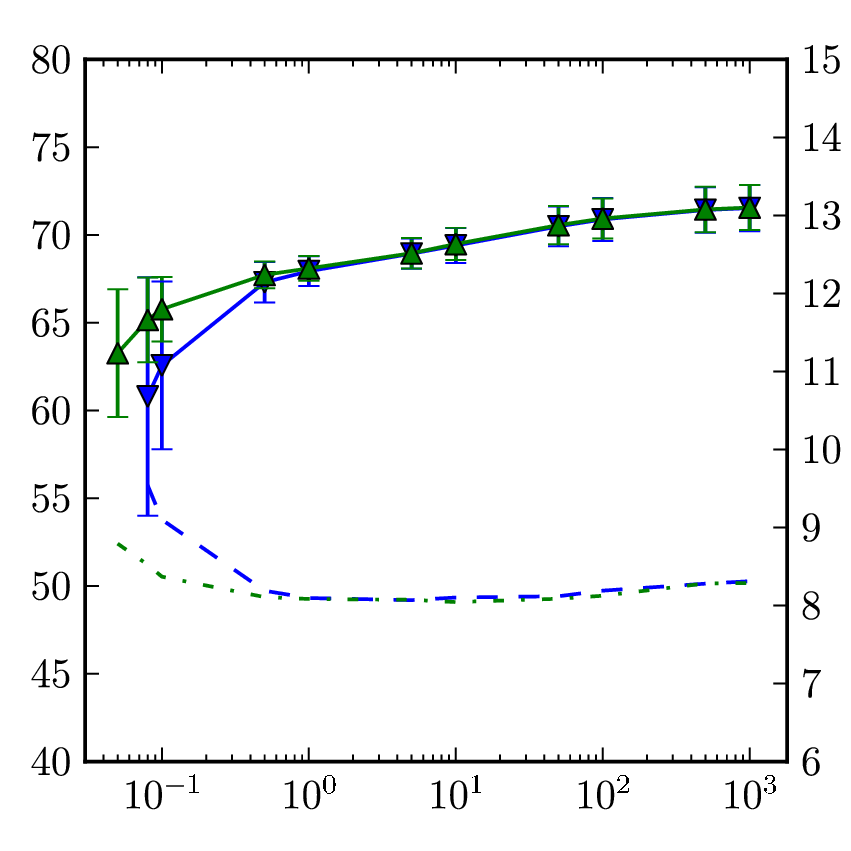}
    \label{fig:trefoil_360_480_vakio_kappa}}
  \subfloat[][$\eta=10$]{\includegraphics[width=0.45\columnwidth]{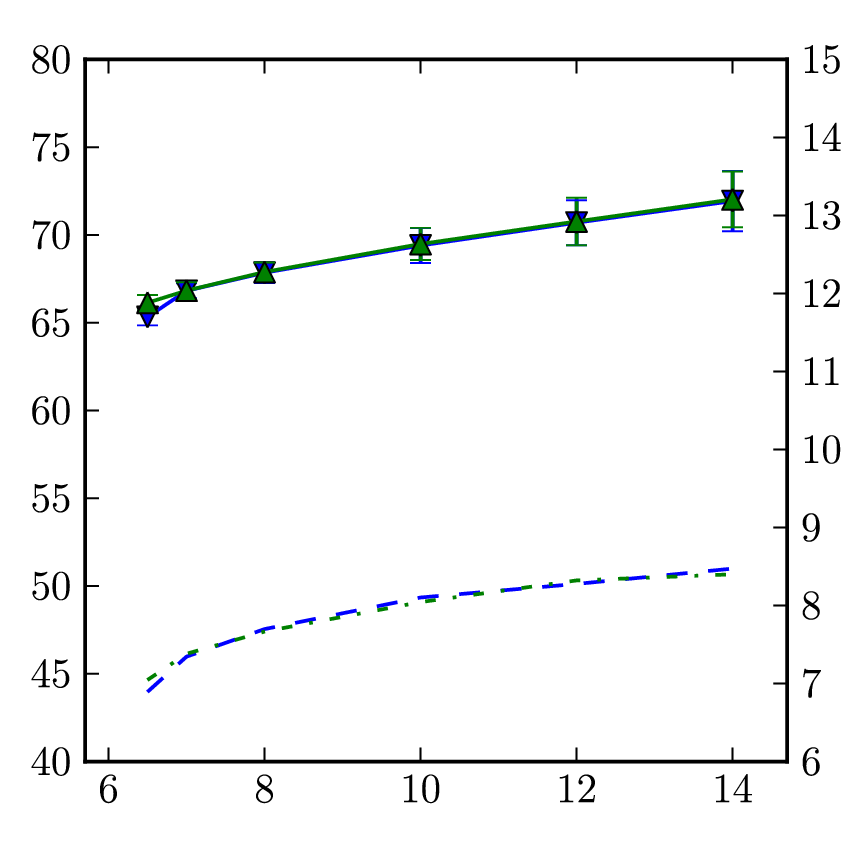}
    \label{fig:trefoil_360_480_vakio_eta}}
  \caption{Total energy and its error limits in normal size and large grid for charge
    $Q=7$ trefoil. Triangles pointing down (solid blue curve) represents $360^3$ and
    triangles pointing up (solid green) $480^3$ lattice, dashed (blue) line the core
    length in first case and dash-dotted (green) in second case. Energy values are on the
    left and core length values on the right $y$-axis.}
  \label{fig:trefoil_360_480}
\end{figure}

Apart from error estimates for each simulation, it is important to note what this
procedure tells about our results in general. Fig.~\ref{fig:trefoil_360_480} depicts the
energies and core lengths of the $H=7$ simulations for $\kappa=10$ (panel
\ref{fig:trefoil_360_480_vakio_kappa}) and for $\eta=10$ (panel
\ref{fig:trefoil_360_480_vakio_eta}) with two different lattice sizes. As can be seen, the
differences are minimal apart from the lowest values of $\eta$, where the energy drops
significantly faster in the smaller lattice. The difference in the soliton itself is not
easily seen until $\eta=0.05$, where we only find a stable soliton in the larger
lattice. This is due to the boundaries exerting pressure on the growing (as $\eta$
decreases) soliton, thus destabilizing it.

The same effect is present at increasing $\kappa$, but here it cannot destabilize the
soliton since increasing $\kappa$ takes us towards a more Faddeev-Skyrme -like system,
where the soliton is stable. The effect, however, is large enough to give rise to fast
decrease in accuracy. This decrease is the reason for the disparity on the ranges covered:
$\eta \in [0.05, 1000]$ and $\kappa \in [6, 14]$ in this work.

As noted in Section~\ref{sec:numerical_methods}, the errors in the core length estimates
are negligible as long as the core is relatively smooth.

\subsection{\label{subsec:soliton_energy}Soliton energy}

Let us now describe in detail the effects of varying the two parameters on the soliton
solutions. First, it should be noted that the energies of the solitons follow the same
$E \propto H^{3/4}$ as those of the Faddeev-Skyrme model. Therefore, in what follows, we
always normalize the energy: $E \rightarrow E/H^{3/4}$.

\begin{figure}[h]
  \centering
  \subfloat[][$\kappa=10$]{\includegraphics[width=0.45\columnwidth]{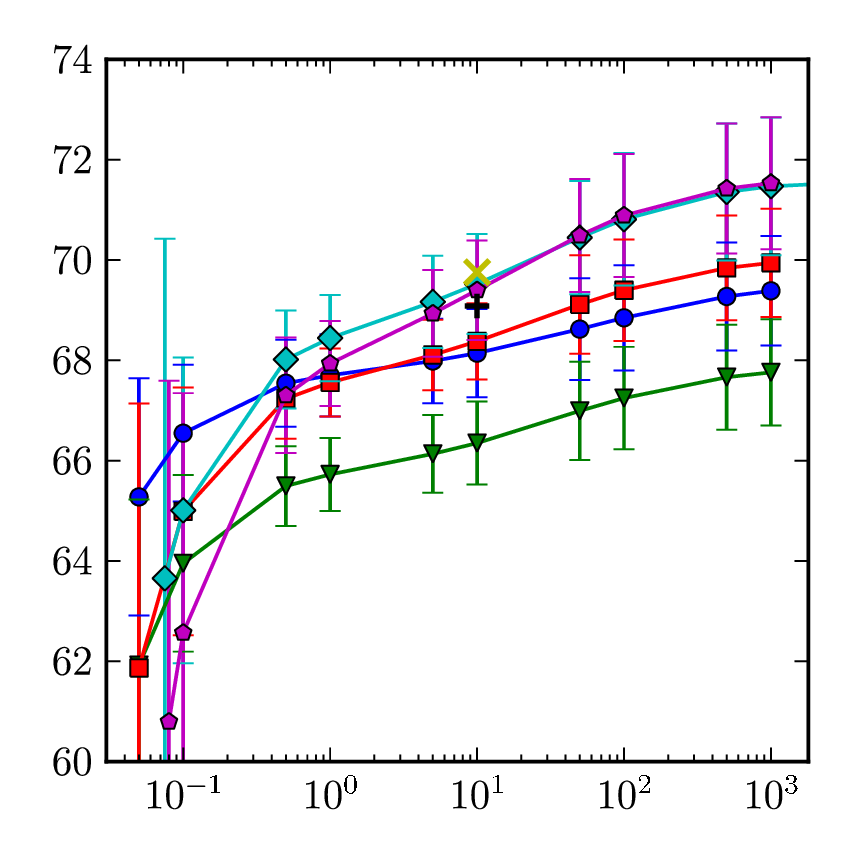}
    \label{fig:energy_vakio_kappa}}
  \subfloat[][$\eta=10$]{\includegraphics[width=0.45\columnwidth]{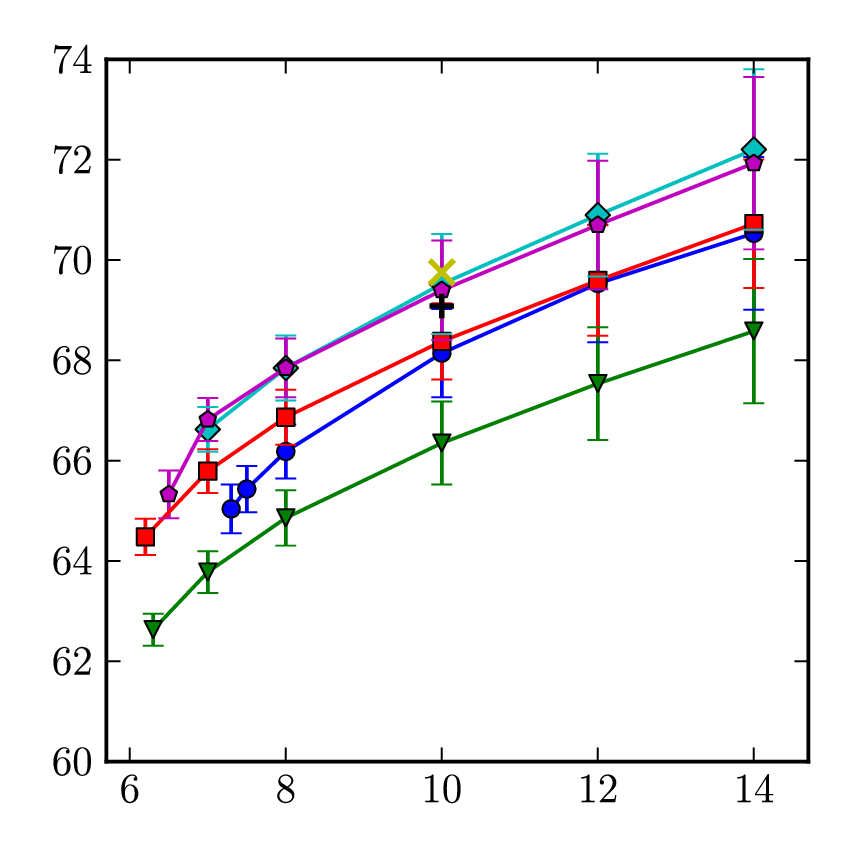}
    \label{fig:energy_vakio_eta}}
  \caption{Plots of the normalized energies ($E/H^{3/4}$) of the minimizers for each $H=1$
    (blue circle), $H=2$ (green triangle), $H=3$ (red square), $H=4$ (cyan diamond) and
    $H=7$ (magenta pentagon). We also include a single data point for $H=5$ (yellow cross)
    and $H=6$ (black plus).}
  \label{fig:energy}
\end{figure}

Because of the asymptotic limit \eqref{eq:EFS}, one expects the energies of the solitons to
approach the limit set by the solitons of the Faddeev-Skyrme model as $\kappa, \eta \to
\infty$. This expectation is found to be true: for each value of the Hopf invariant at
small values of $\kappa$ or $\eta$, the energy is well below the limit and starts growing
towards the value of the Faddeev-Skyrme model when either of the parameters is increased,
as seen in Figure~\ref{fig:energy}, where we plot the energies of all the identified local
minima for all investigated values of $H$ as a function of $\eta$ (panel
\ref{fig:energy_vakio_kappa}) and $\kappa$ (panel \ref{fig:energy_vakio_eta}).

It is worth noting at this point that our program seems to systematically slightly
overestimate the energy: for $H=4$, our values at $\kappa=14,\eta=10$ and
$\kappa=10,\eta=1000$ are about $5\%$ above those reported for the Faddeev-Skyrme model
\cite{Sutcliffe:2007ui}, where an underestimate of about $1\%$ was reported. We believe
that since our results show a clear approach towards an asymptotic value, this is an
acceptable accuracy for such a simple differentiation scheme as used here. The accuracy
could be improved by simply using smaller lattice spacings, but that seems unnecessary for
the purposes of this work and would be computationally too expensive for such a large
number of simulations.

We also note that at values of $\eta=10, \kappa \ge 10$ and $\eta \ge 10, \kappa=10$, the
order of the normalized energies is the same as in the Faddeev-Skyrme model
\cite{Battye:1998zn, Sutcliffe:2007ui, Hietarinta:1998kt, Hietarinta:2000ci}, further
emphasizing the close relationship of the models. It is interesting to note, that the same
order also appears in the extended Faddeev-Skyrme model \cite{2009JHEP...11..124F}.  We
include two extra data points at $\kappa=\eta=10$ for $H=5,6$ to further demonstrate
this. No other simulations were done at $H=5,6$. At lower values of the $\eta$, the order
changes, but simultaneously the error bars grow significantly. For small $\kappa$ the
order would seem to persist, but this is not the case: the boundary of the region of stable
solitons is reached at different values of $\kappa$ for different values of $H$; for
example, as $H=1$ reaches the boundary before $H=2$ it will necessarily have lower energy
below some limiting value of $\kappa$, thus disrupting the order as is explicitly seen to
occur for small $\eta$.

We want to emphasize the fact that the boundary we discover is a bound on the values of
$\kappa,\eta$ \emph{above which} stable solitons can be found. There is no evidence of
their existence below the boundary, but it cannot be ruled out using our methods. Also,
even for our methods, the bound can be pushed slightly downwards by using more accurate
lattices, but instability still eventually occurs (up to what is computationally feasible)
as seen for $H=1$ in \cite{Jaykka:2009rw}.

\subsection{\label{subsec:soliton_core_lenght}Soliton core length}

Next, we turn to the length of the soliton core, i.e. the length of the curve
$\phi^{-1}(0,0,-1) \in \sphere^3 \isom \reals^3 \union \{\infty\}$. 

\begin{figure}[h]
  \centering
  \subfloat[][$\kappa=10$]{\includegraphics[width=0.45\columnwidth]{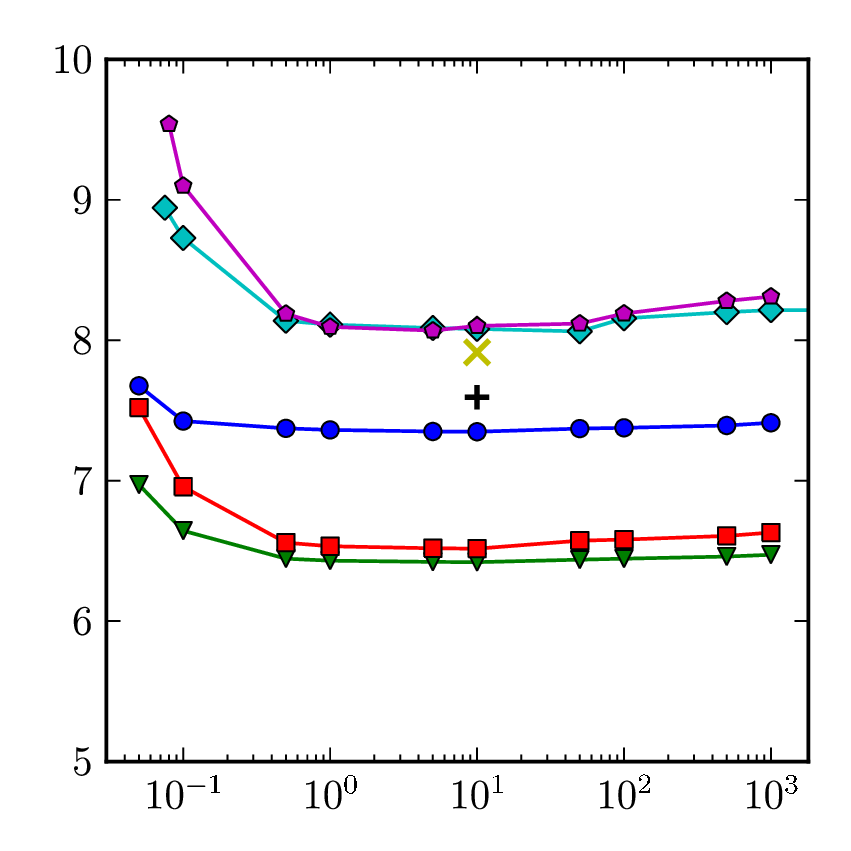}
    \label{fig:corelength_vakio_kappa}}
  \subfloat[][$\eta=10$]{\includegraphics[width=0.45\columnwidth]{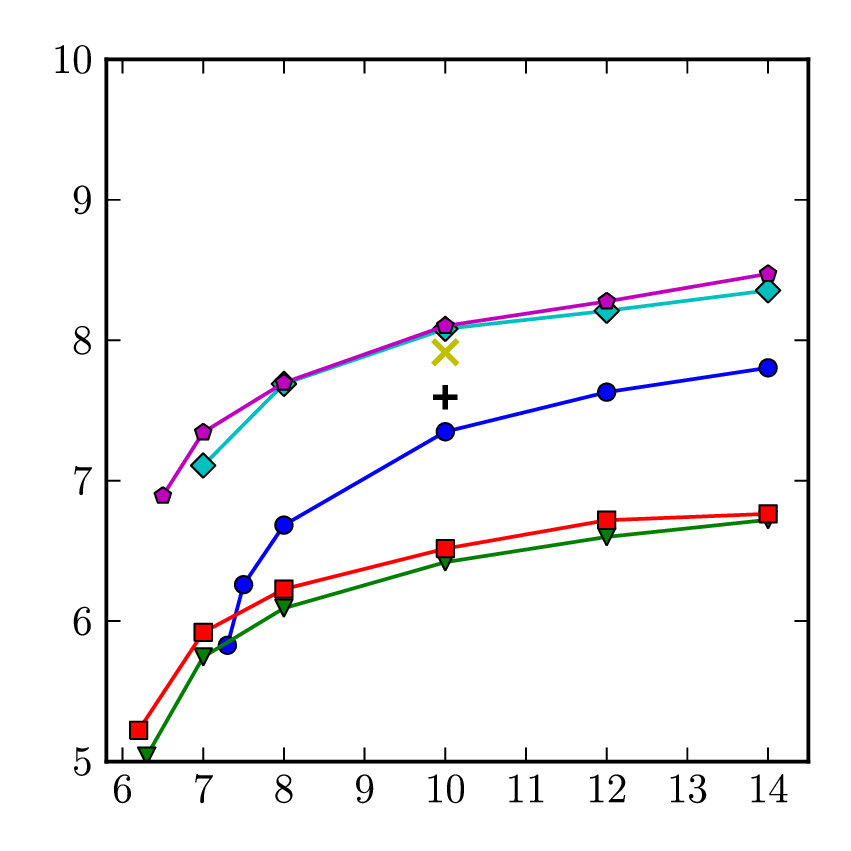}
    \label{fig:corelength_vakio_eta}}
  \caption{Plots of the normalized core lengths ($C/H^{3/4}$) of the minimizers for each
    $H=1$ (blue circle), $H=2$ (green triangle), $H=3$ (red square), $H=4$ (cyan diamond)
    and $H=7$ (magenta pentagon). We also include a single data point for $H=5$ (yellow
    cross) and $H=6$ (black plus).}
  \label{fig:corelength}
\end{figure}

Sutcliffe found that core length for Faddeev-Skyrme Hopfions follow curve $\gamma H^{3/4}$
where his data was fitted to the curve to produce value $\gamma=7.86$
\cite{Sutcliffe:2007ui}. We plot the core lengths of our soliton solutions in
Figure~\ref{fig:corelength}. It is immediately obvious from panel
\ref{fig:corelength_vakio_eta} that the solitons collapse to zero size as $\kappa \to 0$
and as $\kappa$ grows, the soliton sizes approach some asymptotic values.. The behavior
of the core length is more complex in panel \ref{fig:corelength_vakio_kappa}, where
$\kappa$ is held constant. Here the core lengths increase, seemingly without limit, as
$\eta$ decreases, but do not seem to behave monotonically. However, even though the method
used to determine the core length from the computational data is extremely accurate, this
gives no information as to how accurately the numerical core approximates its continuum
counterpart. The seemingly increasing size of the core for large values of $\eta$ falls
within the estimated accuracy of the program, so it can very well be a numerical
artifact. A further evidence in favor of this was received by an additional simulation
performed for $H=4, \kappa=10, \eta=10000$, which gives very slightly shorter core length
as $H=4, \kappa=10, \eta=1000$.

We note that, again, our calculation gives slightly larger values than those reported for
the Faddeev-Skyrme Hopfion in \cite{Sutcliffe:2007ui}. However, this falls well within the
estimated accuracy of the program and the results show a clear approach towards a limiting
value, which is more important than the exact value of that limit.

Because every studied soliton follows the same pattern of decreasing core length with
decreasing $\kappa$, and increasing core length with decreasing $\eta$, we conjecture that
this is a general feature of this model: for core length $C$ one has $\forall
H\in\integers: \; \lim_{\eta \to 0} C(H) = \infty$ and $\lim_{\kappa \to 0} C(H) = 0$. If
this is true, it raises a tantalizing possibility: start with a Faddeev-Skyrme model
Hopfion and then begin decreasing $\kappa$ and $\eta$ in such a way that the collapsing
and exploding effect of their reduction balance. It is unclear whether such a procedure is
possible, but if it is, will it give a stable Hopfion at the limit $\kappa=\eta=0$?  That
is, a Hopf soliton in pure Ginzburg-Landau model. If such a soliton can be constructed, it
seems to require an exact balance between the growing and shrinking effects of decreasing
$\eta$ and $\kappa$, and as such, is probably not possible numerically.

\subsection{\label{subsec:region_supporting_stable_solitons}Region supporting stable solitons}

The stable/unstable border is very difficult to analyze because the existence of
nontrivial local solutions depends on the topological invariant, $\kappa$ and $\eta$, and
their detection depends on the initial state. When the starting point of a simulation is
an old solution instead of the configuration \eqref{eq:Psiwithsech} and
\eqref{eq:initialA}, the stability of the simulation also depends on how much $\kappa$ (or
$\eta$) is changed from the values used to produce the starting point.

This is best illustrated with charge $Q=4$ solutions. The minimizers obtained from an
initial configuration \eqref{eq:Psiwithsech} using parameters $p=1$ and $q=4$ have a
completely different shape than those obtained from $p=q=2$. They also have consistently
higher final energies, but we were able to find stable solutions for smaller values of
$\kappa,\eta$ with the first choice of parameters. This reflects on the fact observed by
Sutcliffe \cite{Sutcliffe:2007ui} that the number of local minima for the Faddeev-Skyrme
model increases with increasing $H$. It appears the same occurs in the present model as
well. It is interesting to note, that the most stable soliton is not always the one with
the lowest energy.

In order to better understand the relationship between the two models, it would be
interesting to see, for each value of $H$, which one of the known Hopfions of the
Faddeev-Skyrme model is most stable in the present model. Our single datapoint in this
would imply that higher value of $E/C$ would sometimes provide a more stable soliton: the
configuration $p,q=1,4$ has higher $E/C$ and can be followed to a lower value of $\kappa$
than $p,q=2,2$. This would give positive support to the conjecture by Babaev that solitons
with longer cores would be more stable \cite{2009PhRvB..79j4506B}.

\section{\label{sec:conclusions}Conclusions}

We have studied the modification of Ginzburg-Landau model proposed by Ward
\cite{Ward:2002vq}. We find that the stable solitons exist for all values of the Hopf
invariant $H$ up to at least $H=7$, but, just like in the situations studied earlier
\cite{Ward:2002vq,Jaykka:2009rw}, this is only possible when the values of the parameters
$\kappa$ and $\eta$ are large enough, and for smaller values, the solitons become unstable
against Derrick-type scaling.

The results suggest a conjecture that in this model, the solitons collapse as $\kappa \to
0$, but expand without limit as $\eta \to 0$. It remains an open question, whether these
effects could be used to precisely balanced each other and provide a way of constructing a
stable knot soliton in the pure Ginzburg-Landau model by starting from a known knot
soliton at the Faddeev-Skyrme limit and reducing the two parameters until $\kappa=0$.

We gain further insight into the relationship between the Ginzburg-Landau and
Faddeev-Skyrme models, by noting that the order of the values of the normalized energies
at different values of the Hopf invariant is the same as in the Faddeev-Skyrme model
\cite{Battye:1998zn, Sutcliffe:2007ui, Hietarinta:1998kt, Hietarinta:2000ci} and extended
Faddeev-Skyrme model \cite{2009JHEP...11..124F}. The fact that in the modified \GL model
the solitons are local minima, not global as in the Faddeev-Skyrme model, is interesting
from condensed matter physics point of view, where local minima are often of great
importance.

\begin{acknowledgments}
  The authors with to thank R.~S.~Ward, E.~Babaev and J.~Hietarinta for useful comments and
  discussions. This work has supported by the Academy of Finland (project 123311) and the
  UK Engineering and Physical Sciences Research Council. The authors acknowledge the
  generous computing resources of CSC~--~IT Center for Science Ltd, which provided the
  supercomputers used in this work.
\end{acknowledgments}

\bibliography{bibliography}

\end{document}